\documentclass[10 pt,twoside]{memoir}
\usepackage[T2A]{fontenc}
\usepackage[utf8]{inputenc}
\usepackage{graphicx}
\usepackage{amsmath}
\usepackage{latexsym, amssymb}
\usepackage[font={small}]{caption} 
\counterwithout{figure}{chapter}
\usepackage[shortlabels]{enumitem}
\setlist{ noitemsep }
\usepackage[symbol]{footmisc}
\usepackage{url}
\captiontitlefont{\small} 
\captionnamefont{\small} 
\usepackage[text={6.5in,9in},centering]{geometry}
\usepackage{microtype} 
\makeatletter
\def\MT@register@subst@font{\MT@exp@one@n\MT@in@clist\font@name\MT@font@list
\ifMT@inlist@\else\xdef\MT@font@list{\MT@font@list\font@name,}\fi}
\makeatother
\makepagestyle{ruled} 
\makeevenfoot{ruled}{}{}{\thepage}
\makeoddfoot{ruled}{}{}{\thepage} 
\makeevenhead{ruled}{}{\leftmark}{} 
\makeoddhead{ruled}{}{\rightmark}{}
\begin{document}
\thispagestyle{empty}
\renewcommand*{\thefootnote}{\fnsymbol{footnote}}

\normalsize

\begin{center}
\Large GRAVITATIONAL MACHINES\footnote[1]{Extracted and \LaTeX'd (with new footnotes) for the arXiv with permission of the author from \emph{Interstellar Communication: A Collection of Reprints and Original Contributions}, A.\,G.\,W.\,Cameron, editor; copyright 1963, W.\,A.\,Benjamin Co., New York, pp.\,115--120. The article was written for the 1962 prize competition of the Gravity Research Foundation (it won fourth place); with gratitude to George Rideout, President, for his encouragement to post the article.}\\
\end{center}

\normalsize
\begin{center}
\textsc{Freeman J. Dyson}\rule[-1.5ex]{0pt}{1.5ex}\\
\footnotesize\textsc{Institute for Advanced Study,}\\
\textsc{Princeton, N.\,J.}\\
\end{center}
\normalsize
\setcounter{footnote}{0}
\renewcommand*{\thefootnote}{\arabic{footnote}}

\begin{quote}\small A gravitational machine is defined as an arrangement of gravitating masses from which useful energy can be extracted. It is shown that such machines may exist if the masses are of normal astronomical size. A simple example of a gravitational machine, consisting of a double star with smaller masses orbiting around it, is described. It is shown that an efficient gravitational machine will also be an emitter of gravitational radiation. The emitted radiation sets a limit on the possible performance of gravitational machines, and also provides us with a possible means for detecting such machines if they exist.
\end{quote}
\normalsize

\noindent The difficulty in building machines to harness the energy of the gravitational field is entirely one of scale. Gravitational forces between objects of a size that we can manipulate are so absurdly weak that they
can scarcely be measured, let alone exploited. To yield a useful output of energy, any gravitational machine must be built on a scale that is literally astronomical. It is nevertheless worthwhile to think about
gravitational machines, for two reasons. First, if our species continues to expand its population and its technology at an exponential rate, there may come a time in the remote future when engineering on an astronomical scale will be both feasible and necessary. Second, if we are searching for signs of technologically advanced life already existing elsewhere in the universe, it is useful to consider what kinds of observable phenomena a really advanced technology might be capable of producing.

The following simple device illustrates the principle that would make possible a useful gravitational machine (see Figure 1). A double star has two components $A$ and $B$, each of mass $M$, revolving around each other in a circular orbit of radius $R$. The velocity of each star is
\begin{equation}
V = (GM/4R)^{1/2}
\end{equation}
where
\begin{equation}
G = 6.7 \times 10^{-8}\,\text{cm}^{3}/\text{sec}^{2}\text{g}
\end{equation}
is the gravitational constant. The exploiters of the device are living on a planet or vehicle $P$ which circles around the double star at a distance much greater than $R$. They propel a small mass $C$ into an orbit which falls toward the double star, starting from $P$ with a small velocity. The orbit of $C$ is computed in such a way that it makes a
close approach to $B$ at a time when $B$ is moving in a direction opposite to the direction of arrival of $C$. The mass $C$ then swings around $B$ and escapes with greatly increased velocity. The effect is almost as if the light mass $C$ had made an elastic collision with the moving heavy mass $B$.
\begin{figure}[h]
\centering\scalebox{0.6}{\includegraphics{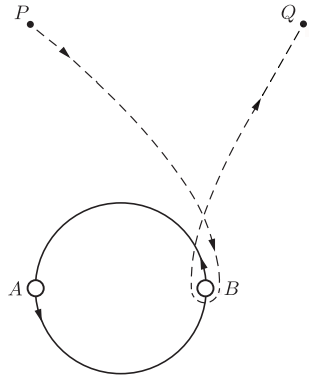}}
\caption{\emph{The solid line indicates the orbit of $A$ and $B$; the dashed line indicates the orbit of $C$.}}
\end{figure}
The mass $C$ will arrive at a distant point $Q$ with velocity somewhat greater than $2V$. At $Q$ the mass $C$ may be intercepted and
its kinetic energy converted into useful form. Alternatively the device may be used as a propulsion system, in which case $C$ merely proceeds with velocity $2V$ to its destination. The destination might be a similar device situated very far away, which brings $C$ to rest by the same mechanism working in reverse.

It is easy to imagine this device converted into a continuously operating machine by arranging a whole ring of starting points $P$ and end points $Q$ around the double star, masses $C$ being dropped inward and emerging outward with increased velocity in a continuous stream. The energy source of the machine is the gravitational potential between the stars $A$ and $B$. As the machine continues to operate, the stars $A$ and $B$ will gradually be drawn closer together, their negative potential energy will increase and their orbital velocity $V$ will also increase. The machine will continue to extract energy from their mutual attraction until they come so close together that orbits passing between them are impossible. For a rough estimate one may suppose that the machine can operate until the distance between the centers of the two stars is equal to 4$a$, where $a$ is the radius of each star. The total energy extracted by the machine from the gravitational field is
then
\begin{equation}
E = GM^{2}\!/8a
\end{equation}
\indent If $A$ and $B$ are ordinary stars like the sun, the radius $a$ is of the order of $10^{11}$ cm. The energy $E$ is then equal to the luminous energy radiated by the stars in a few million years. Under these conditions the available gravitational energy may be exploited, but it is of minor importance compared with the luminous energy of the system. A technically advanced species would presumably put its main efforts
into harnessing the luminous energy.

If the stars $A$ and $B$ are typical white dwarfs, the situation is entirely reversed. In that case the optical luminosity is less than that of the sun by a factor of about a thousand, while the available gravitational energy is increased by a factor of a hundred. It is therefore logical to expect that around a white-dwarf binary star a technology based on gravitational energy might flourish. For purposes of illustration, let us assume
\begin{align}
M &= \mbox{1 solar mass} = 2 \times 10^{33}\,\text{g}\\
a &= 10^{9}\,\text{cm}\\
R &= 2a = 2 \times 10^{9}\,\text{cm}
\end{align}
Then we find
\begin{align}
V &= 1.3 \times 10^{8}\,\text{cm/sec}\\
E &= 3\times 10^{49}\,\text{ergs}
\end{align}
The orbital period of the binary star is 
\begin{equation}
P = 100\,\text{sec}
\end{equation}
A search for eclipsing binaries of such short period among the known white dwarfs was suggested many years ago by H.\ N.\ Russell [1]. The search was subsequently made by F.\ Lenouvel [2], with negative results.
The negative result is not surprising, since the total number of identified white dwarfs is very small.

A white-dwarf binary star with the parameters (4) to (9) would have the interesting property that it could accelerate delicate and fragile objects to a velocity of 2000 km/sec at an acceleration of 10,000\,$g$, without doing any damage to the objects and without expending any rocket propellant. The only internal forces acting on the accelerated objects would be tidal stresses produced by the gradients of the gravitational fields. lf the over-all diameter of the object is $d$, the maximum tidal acceleration will be of the order of
\begin{equation}
A = GMd/a^{3} = \tfrac{1}{8}d
\end{equation}
For example, if $A$ is taken to be 1 earth gravity, then $d = 80\,\text{meters}$. So a large space ship with human passengers and normal mechanical construction could easily survive the 10,000\,$g$ acceleration. lt may be imagined that a highly developed technological species might use white-dwarf binaries scattered around the galaxy as relay stations for heavy long-distance freight transportation.

An important side effect of a short-period white-dwarf binary would be its enormous output of gravitational radiation. According to the standard theory of gravitational radiation [3], which is not universally accepted, a pair of stars of equal mass moving in a circular orbit radiates gravitational energy at a rate
\begin{equation}
W= 128V^{10}\!/\hspace*{1pt}5Gc^{5}
\end{equation}
where $c$ is the velocity of light. It would be extremely valuable if we could observe this radiation, both to verify the validity of the theoretical formula (11) and to detect the existence of white-dwarf binaries. Inserting the value of $V$ from (7) into (11), we find
\begin{equation}
W= 2 \times 10^{37}\,\text{ergs/sec}
\end{equation}
which is 5000 times the sun's optical luminosity. Comparing (12) with (8), we see that the gravitational radiation itself will limit the lifetime of this white-dwarf binary to about 40,000 years. However, since the dependence of $W$ on $V$ is so extreme, a binary with $V = 5 \times 10^{7}\,\text{cm/sec}$ could live for many millions of years. A technologically advanced species might then choose the value of $V$ to suit its particular purposes.

Assuming the value (12) for the intensity of a source of gravitational waves at a distance of 100 parsecs, we find that the signal to be detected on earth has the intensity
\begin{equation}
I= 2 \times 10^{-5}\,\text{erg/cm$^{2}$-sec}
\end{equation}
Unfortunately the period of the radiation is 100 sec, which is not short enough to be observed with the existing apparatus of J.\,Weber [4].
However, it is quite likely that a detector could be built which would be sensitive to the incident flux (13) at a period of 100 sec. This would then allow us to detect by its gravitational radiation any white-dwarf binary of period 100 sec within a volume of space containing over half a million stars [5].

According to astrophysical theory [6], a white dwarf is not the most condensed type of star that is possible. A still more condensed form of matter could exist in ``neutron stars,'' which would have masses of the same order as the sun compressed into radii of the order of 10 km. Whether neutron stars actually exist is uncertain; they would be very faint objects, and the fact that none has yet been observed does not argue strongly against their existence.

If a close binary system should ever be formed from a pair of neutron stars, the consequences would be very interesting indeed. Consider for example a pair of stars of solar mass, each having radius
\begin{equation}
a = 10^{6}\,\text{cm}
\end{equation}
and with their centers separated by a distance $2R = 4a$. According to (1) and (11), each star moves with velocity
\begin{equation}
V = 4 \times 10^{9}\,\text{cm/sec}
\end{equation}
in an orbit with period 5 milliseconds, and the output of gravitational radiation is 
\begin{equation}
W = 2 \times 10^{52}\,\text{ergs/sec}
\end{equation}
But by (3), the gravitational energy of the pair is at this moment
\begin{equation}
E = 3 \times 10^{52}\,\text{ergs}
\end{equation}
Thus the whole of the gravitational energy is radiated away in a violent pulse of radiation lasting less than 2 sec. A neutron-star binary beginning at a greater separation $R$ will have a longer lifetime, but the final end will be the same. According to (11), the loss of energy by gravitational radiation will bring the two stars closer with
ever-increasing speed, until in the last second of their lives they plunge together and release a gravitational flash at a frequency of about 200 cycles and of unimaginable intensity.

A pulse of gravitational radiation of magnitude (17) at a frequency around 200 cycles should be detectable with Weber's\footnote{At the time of this article's writing, Joseph Weber was engaged in the world's first attempt to
detect gravitational waves, using a detector of modest size and cost that he constructed himself. Other researchers built similar detectors but failed to replicate his results, and Weber's claims of real signals were judged by the community to be instrumental noise. James L.\ Levine and Richard L.\ Garwin, \emph{Phys.\,Rev.\,Lett.} \textbf{31} (1973) 173-176, 176-180; \emph{Phys.\,Rev.\,Lett.} \textbf{33} (1974) 794-797. Announcing a detection of gravitational waves, the LIGO team cited Weber's pioneering work: B.\ P.\ Abbott \emph{et al.},  \emph{Phys.\,Rev.\,Lett.} \textbf{116}, 061102 (2016).} existing equipment [4] at a distance of the order of 100 Mparsecs. So the death cry of a binary neutron star could be heard on earth, if it happened once in 10 million galaxies. It would seem worthwhile to maintain a watch for events of this kind, using Weber's equipment or some suitable modification of it.\footnote{Such a signal, GW170817, \emph{was} detected by the LIGO team: B.\ P.\ Abbott \emph{et al.},  \emph{Phys.\,Rev.\,Lett.} \textbf{119}, 161101 (2017).}

Clearly the immense loss of energy by gravitational radiation is an obstacle to the efficient use of neutron stars as gravitational machines. It may be that this sets a natural limit of about $10^{8}\,\text{cm/sec}$ to the velocities that can be handled conveniently in a gravitational technology. However, it would be surprising if a technologically advanced species could not find a way to design a nonradiating gravitational machine, and so to exploit the much higher velocities which neutron stars in principle make possible.

In conclusion, it may be said that the dynamics of stellar systems, under conditions in which gravitational radiation is important, is a greatly neglected field of study. In any search for evidences of technologically advanced societies in the universe, an investigation of anomalously intense sources of gravitational radiation ought to be included.\\

{\centering
\textbf{References}

}

\small

\begin{enumerate}[label={[\arabic*]}]
\item H.\ N.\ Russell, Centennial Symposia, \emph{Harvard Monographs,} \textbf{7}, 187 (1948).
\item F.\ Lenouvel, \emph{J.\ Observateurs,} \textbf{40}, 15 (1957).
\item L.\ Landau and E.\ Lifshitz, \emph{Classical Theory of Fields}, trans.\ by M.\ Hamermesh, Addison-Wesley, Reading, Mass., 1961, p.\,332.
\item J.\ Weber, \emph{Phys.\ Rev.,} \textbf{117}, 306 (1960).
\item R.\ P.\ Kraft, J.\ Matthews, and J.\ L.\ Greenstein have discussed the extremely interesting object Nova WZ Sagittae, a binary star with one white-dwarf component and an observed orbital velocity of about $7 \times 10^{7}\,\text{cm/sec}$. They point out that gravitational radiation from this object is certainly important and may be detectable. The author is indebted to Dr.\ Kraft for a preprint of this work.
\item J.\ R.\ Oppenheimer and G.\ M.\ Volkoff, \emph{Phys.\ Rev.,} \textbf{55}, 374 (1939); A.\ G.\ W.\ Cameron, \emph{Astrophys.\ J.,} \textbf{130}, 884 (1959).
\end{enumerate}
\vspace*{0.6in}

\footnotesize

\noindent \emph{Typist's note}

After the detection of the gravitational wave GW170817, Jason T. Wright (\emph{Physics Today}, \textbf{72}, 5, 12, 2019) reminded the community that many of its features had been predicted by Dyson more than half a century earlier. Dyson's article was published only once, in Cameron's long out of print collection, though a scan may be found at the web site of the Gravity Research Foundation (\url{https://www.gravityresearchfoundation.org}). Dyson thought it had been reprinted (in his \emph{Selected Papers}, AMS Press, 1996, forward by Elliot H. Lieb) but it was not. Hoping to make the article easier to find, I wrote Dyson for his permission to post it at the arXiv. He happily consented to this proposal, but wanted to add a new footnote about J.\,Weber's heroic research (he provided the text). A second footnote was added to cite the experimental confirmation of his predictions. Finally, the abstract did not appear in the Cameron printing, but did (titled ``Summary'') in the original submission to the Gravity Research Foundation, and it has been taken from that.

\hfill D.\,Derbes

\end{document}